\newif\ifAbstract
\Abstractfalse

\newif\ifProceeding
\Proceedingfalse

\newif\ifJournal
\Journalfalse

\newif\ifSaveSpace
\SaveSpacefalse

\newif\ifUsePDF
\UsePDFfalse

\newif\ifSkip
\Skipfalse

\Proceedingtrue

\ifJournal
\else\ifProceeding
  \documentclass[twoside,11pt]{article}
  \usepackage{epsfig,color,makeidx,url}

  \usepackage{times}
  \usepackage{mathptm}

\fi
\fi

\title{The Geometric Thickness of Low Degree Graphs}
\author {
\sc Christian A.~Duncan
\\\normalsize Department of Computer Science
\\\normalsize University of Miami
\\\normalsize \tt duncan@cs.miami.edu
\and
\sc David Eppstein\thanks{ This work is supported in part by 
the NSF under grant CCR-9912338.}
\\\normalsize Department of Computer Science
\\\normalsize University of California, Irvine
\\\normalsize \tt eppstein@uci.edu
\and 
\sc Stephen G. Kobourov\thanks{ This work is supported in part by 
the NSF under grant ACR-0222920.}
\\\normalsize Department of Computer Science
\\\normalsize University of Arizona
\\\normalsize \tt kobourov@cs.arizona.edu
}

\newtheorem{theorem}{Theorem}[section]
\newtheorem{lemma}[theorem]{Lemma}

\newcommand{\qed}{\rule{6pt}{6pt}}
\newcommand{\bang}{$\bf\Rightarrow\Leftarrow$}
\newenvironment{proof}{\noindent{\it Proof:}\hspace*{1em}}{\qed \medskip}
\newenvironment{contradiction-proof}{\noindent{\it Proof:}\hspace*{1em}}{\bang\ \qed \medskip}

\newcommand{\HalfSpace}{\edef\baselinestretch{0.9}\Large\normalsize}

\newcommand{\Comment}[1]{\relax}

\begin{document}

\date{}
\maketitle
\pagestyle{myheadings}
\markboth{}{}


\begin{abstract}
We prove that the geometric  thickness of graphs whose maximum  degree
is no more than  four  is two.  All of   our algorithms run in  $O(n)$
time,  where $n$ is  the  number of vertices   in  the graph.  In  our
proofs,  we present an  embedding  algorithm for  graphs  with maximum
degree three  that uses  an $n  \times  n$  grid  and a   more complex
algorithm for  embedding a graph  with maximum  degree  four.  We also
show a variation using orthogonal edges for maximum degree four graphs
that also uses an $n \times n$ grid.  The results have implications in
graph theory, graph drawing, and VLSI design.

\end{abstract}

\section{Introduction}
\label{sec::introduction}
The {\em thickness} of a graph $G=(V,E)$ is the smallest number of
planar subgraphs needed to decompose $G$.  That is, we wish to find
the smallest number $i$ such that we can partition the edges $E$ into
$i$ different planar subgraphs; see~\cite{mutzelthickness98} for a
survey.  Using the terminology from VLSI, these subgraphs are referred
to as the {\em layers} of the original graph.
In several applications, including VLSI layouts~\cite{AggKlaSho90},
visualization of software development~\cite{ck-sgbv-03}, and graph
drawing~\cite{eknl-sgd-03} it is necessary that the vertex locations
be consistent across the layers. It is well known that a planar graph
can be drawn in the plane without crossings, using arbitrary placement
of the vertices and Jordan curves representing the edges. However, the
complexity of the edges can be quite large; if the edges are
represented by non-crossing polygonal curves, then $O(n)$ bends per
edge are needed~\cite{pach98embedding}.

If we add the requirement that edges on all layers must be
represented by straight-line segments, we arrive at the notion of {\em
geometric thickness}. Geometric thickness
requires that the vertices for each subgraph's embedding be in the
same location and that each edge be drawn with a straight-line
segment~\cite{deh-gtcg}.
In a recent paper on geometric thickness~\cite{e-stgt-02}, one of the authors
posed as an open problem bounding the geometric thickness of a graph as
a function of its degree. In this paper we show that graphs of maximum
degree three and four have geometric thickness two. The underlying
algorithms are efficient and easy to implement. We have implemented
the degree-three algorithm, and Figure~\ref{fig:coxeter} shows 
two-layer drawings produced by our implementation.
   
\subsection{Related Work}
Our work is related to results on graph thickness, geometric
thickness, rectangle visibility, graph arboricity and simultaneous
embeddings.
 
Since Kainen's work on graph thickness~\cite{kainen-thickness73} there
has been a great deal of work on graph thickness and several
restrictions, such as geometric thickness~\cite{deh-gtcg} and book
thickness~\cite{bk-btg-79}.  Recent results include the asymptotic
non-equivalence of graph thickness and geometric
thickness~\cite{e-stgt-02} as well the asymptotic non-equivalence of
geometric thickness and book
thickness~\cite{oai:arXiv.org:math/0109195}.  Geometric thickness on
the grid has also been considered. In~\cite{wood-grid03} Wood shows
that the vertices of an $n$-vertex $m$-edge graph can be positioned in
a $\lceil\sqrt{n}\rceil\times \lceil\sqrt{n}\rceil$ grid and the edges
assigned to $O(\sqrt{m})$ layers, so that each edge is drawn with at
most one bend and no two edges on the same layer cross.

Our result for graphs of degree three is obtained by combining bounds on {\em linear arboricity} with known {\em simultaneous embedding} techniques.
The linear arboricity of a graph is the minimum number of disjoint unions of paths needed to cover the edges of the graph.
Akiyama {\em et al.}~\cite{lncs108*38,MR82k:05085a,MR82k:05085b} show
that the linear arboricity of cubic graphs is two and Alon {\em et
al.}~\cite{MR2002d:05068} present upper bounds on the linear arboricity
of regular graphs.
Two planar graphs on the same vertex
set, $G_1=(V,E_1)$ and $G_2=(V,E_2)$ can be simultaneously embedded if
there exists a point set $P$ in the plane for the vertices in $V$ such
that each of $G_1$ and $G_2$ can be drawn on $P$ with straight-line edges
and no crossings.  Brass {\em et
al.}~\cite{WADS::Brass2003} show that pairs of paths, cycles, and
caterpillars can be simultaneously embedded on the $O(n)\times O(n)$
grid.
For our degree four results, we extend these results to graphs formed by disjoint unions of cycles.

Related results have also been obtained in rectangle visibility
problems. In particular Bose {\em et al.}~\cite{bose96rectangle} show
that any graph with maximum degree four is a rectangle visibility
graph.

\subsection{Our results}

In this paper we assume that our input graph is connected. 
Otherwise, we could simply embed each of the connected components separately and
then combine the resulting embeddings.  

We begin with a simple argument and an algorithm for the maximum
degree three case. We show that graphs with maximum degree three can
be decomposed into two layers such that the vertices in both layers
are in the same position and the edges are straight-line
segments. Thus, such graphs have geometric thickness two. We present
and implement an
$O(n)$ time algorithm for the decomposition, where $n$ is the number
of vertices in the graph. Moreover, 
the layout requires $n \times n$ grid area.

Next we generalize the technique to the degree four case.
Our main result is that graphs with maximum
degree four have geometric thickness two. We present a linear time algorithm for the
decomposition of the graph into two layers and for finding a pointset
$P$ that realizes the graph in two layers. 
This part of our paper is based on a novel simultaneous embedding theorem:
any two graphs each with maximum degree two have a simultaneous embedding.
The resulting graph requires a grid with large area.
Consequently, we also introduce
a variant of our problem using orthogonal edges with only one bend per edge.
This allows us to easily compute an orthogonal two-layer
embedding in linear time on an $n \times n$ grid.

\section{Degree Three Graphs}
\label{sec::degreeThree}
We will show that graphs of maximum degree three have geometric
thickness two in two steps: (1) decomposing the graph into two
subgraphs and (2) showing that the resulting two subgraphs can be
simultaneously embedded.

The first step uses a known result on graph arboricity.  A {\em linear
forest} is a forest composed only of paths.  A graph $G$ has {\em
linear arboricity} $l$ if the edges of $G$ can be partitioned into $l$
linear forests.  This definition is purely combinatorial and
does not require any embedding of the graph.
Assume that we have a connected, maximum-degree-three graph $G
= (V,E)$, that is for every $v \in V$, $d(v) \leq 3$.
In~\cite{MR82k:05085a,MR82k:05085b}, Akiyama {\em et al.} prove that
the linear arboricity of $G$ is at most two; see
also~\cite{lncs108*38,MR89m:05086,MR2002d:05068}.  This alone does not
prove that the graphs have geometric thickness two but only that they
have graph thickness two.

To implement the decomposition of Akiyama {\em et al.}, we use a simple DFS technique:

\begin{lemma}
A graph of degree at most three can be decomposed into two linear forests in linear time.
\end{lemma}

\begin{proof}
We start the decomposition by assigning the edges of a DFS tree to the two layers according to the parity of the distance from the root to the edge: edges incident to the root are assigned to the first layer, edges incident to children of the root are assigned to the second layer, and so on.
We view the DFS tree as oriented downwards, from the root at the top of the tree.
We next assign non-tree edges to layers, one at a time according to the DFS order of
their bottom endpoints, so that no vertex has all three edges belonging to the same layer
(but allowing edges in a layer to form cycles that we will later repair).
When possible, an edge is assigned to a different layer from its top endpoint's
downward-going tree edges; edges assigned in this way can form at most one cycle with
the other edges of the same layer because such a cycle could have no topmost vertex except
at the tree root where there may be two incident non-tree edges.
If two non-tree edges exist at a vertex, and one leads to the tree root, that one is assigned first.
Any remaining unassigned edges occur when a vertex has an incoming
tree edge and assigned non-tree edge in the same layer, precluding the third edge at that vertex
from being assigned to that layer.  We assign these edges to the other layer,
which is always possible because the top endpoint of the edge
must have incoming and outgoing tree edges in different layers.
Edges assigned in this way cannot be part of single-layer cycles because their bottom endpoint
only has one edge of the assigned color.

After this assignment process, the two layers together contain at most a single cycle passing through the root of the DFS tree.  To remove this cycle, we change the assignment of one of the two cycle edges incident to the DFS root.  This change of assignment will not create a cycle in the other layer, unless the chosen edge is also incident to the endpoint of the path in the other layer that starts from the DFS root.  Therefore at least one of the two edges can be safely reassigned creating the desired decomposition.
\end{proof}

\begin{figure}
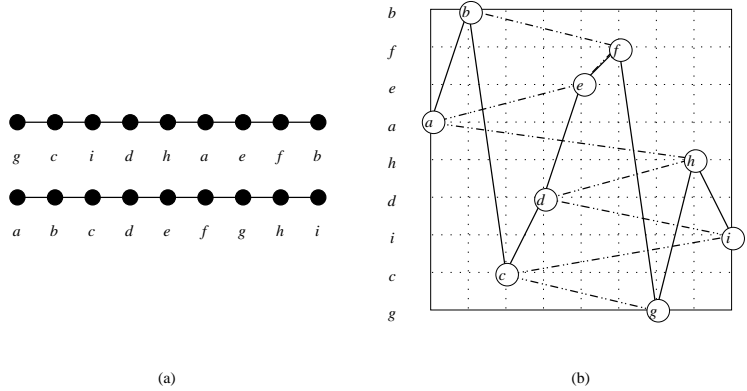

\begin{center}
\ifUsePDF
\includegraphics{two-paths.pdf}
\else
\input{two-paths.pstex_t}
\fi
\end{center}
\vspace{-.5cm}
\caption{\sf\small Simultaneous embedding of two paths.  (a) The two paths: each vertex appears twice, once per path.
(b) The embedding of both paths: the solid line represents one path and the dashed line represents the other path.
}
\label{fig:two-paths}
\end{figure}

For the second step of our geometric thickness two layout,
we augment each of the two edge sets, by adding artificial edges,
until each set is a single path that visits all the
nodes. Using an algorithm from Brass {\em et
al}~\cite{WADS::Brass2003} we can simultaneously embed these two
paths. Since we will generalize this technique in the next section, we
present an overview of the algorithm here. Assume we have two paths
formed by $n$ vertices; see Figure~\ref{fig:two-paths}.  These paths
present two orderings of the vertices.  For a vertex $v$, let
$\sigma_1(v)$ and $\sigma_2(v)$ be $v$'s position in the first and
second paths, respectively.  We create an $n \times n$ grid and, for
each vertex $v$, we place $v$ at grid position $(\sigma_1(v),
\sigma_2(v))$.  That is, the $x$ position is determined by the
vertex's ordering in the first path and the $y$ position by the
ordering in the second path. Note that by the definition of the
placement of the vertices, the first path has strictly increasing $x$
coordinates and the second path has strictly increasing $y$
coordinates. Thus, each path by itself is non-crossing.
We then remove the artificial edges from our drawing.

Since a maximum degree three graph can be decomposed into a union of
two linear forests and these forests can be augmented to two paths,
the theorem below follows:

\begin{theorem}
Graphs with maximum degree three have geometric thickness at most two.
A geometric thickness two embedding of a degree three graph onto
an $n\times n$ grid can be found in time $O(n)$.
\end{theorem}

Examples of graph drawings created by this algorithm are depicted in Figure~\ref{fig:coxeter}.

\section{Degree Four Graphs}
\label{sec::degreeFour}
Suppose that we have a maximum-degree-four graph $G = (V,E)$. 
That is, for every $v \in V$, $d(v) \leq 4$.
Since every graph with only even-degree vertices 
has an Euler tour (see, for example,~\cite{MR96i:05001}),
we can prove the following lemma, similar to linear arboricity except that 
some paths may be closed into cycles:

\begin{lemma}
Any graph of maximum degree four can be partitioned into two 
subgraphs each of which is a disjoint union of cycles and paths.
\end{lemma}

\begin{proof}
Since every graph of maximum degree four is a subgraph of some four-regular graph
with added vertices and edges and since every four-regular graph has an Euler
tour,
let $T$ be the tour associated with this four-regular graph.
Let our edges be numbered $e_0, e_1, \ldots$ according to their ordering in 
the tour starting with any edge.
We now partition our edges into two sets, the even-numbered edges and the 
odd-numbered edges, forming two subgraphs $G_0$ and $G_1$ respectively.

We claim that $G_0$ and symmetrically $G_1$ are each disjoint forests of 
cycles and paths.
Since $G_0$ consists of only even-numbered edges, all vertices in $G_0$ have
degree two.
The same applies for $G_1$.
Next we remove the edges and vertices
that are in the four-regular graph but not $G$.
This may introduce vertices 
of degree zero, but we can simply remove these vertices
from the subgraph or treat them as zero-length paths.
\end{proof}

Notice that the two subgraphs are not vertex disjoint from each other as they 
certainly share vertices in common.
We now begin the task of embedding the two subgraphs simultaneously, using an 
enhancement to the technique
of embedding two cycles presented in~\cite{WADS::Brass2003}.
Without loss of generality, assume the union of the two graphs is connected,
if it weren't we could handle each component separately.
As mentioned earlier, since the subgraphs have maximum degree two, they are 
collections
of paths and cycles.
We can assume at least one of $A$ and $B$ is disconnected; otherwise
we could simply apply the technique from~\cite{WADS::Brass2003} to embed two cycles.
To simplify our arguments, let us connect all paths in $A$ (similarly $B$) 
into a single cycle by adding some temporary edges.
This makes $A$ and $B$ forests of cycles, with a slight exception if there is only 
one path consisting of one edge; however, as will become quite apparent this 
poses no problem for the algorithm.

We shall provide an ordering of the individual cycles of $A$ and $B$ 
and label them $c_0, c_1, \ldots, c_k$.
We embed all cycles on a grid so that a cycle $c_i$ from $A$ has $x$-coordinates
forming a consecutive subinterval of the range $0$ to $n-1$ and $y$-coordinates
scattered over the entire range from $0$ to $n-1$.
Similarly, a cycle $c_i$ from $B$ has $y$-coordinates forming a consecutive
subinterval and $x$-coordinates scattered over the entire range.
We also guarantee that no two vertices share the same $x$ or $y$ coordinate.

Once the cycles have been ordered and their initial vertices have been determined,
the union of all the cycles in $A$
forms an ordering of the $x$-coordinates from 0 to $n-1$, and in $B$ the union
forms an ordering of the $y$-coordinates.
Also, each cycle $c_i$, will have an initial vertex, $v_i$.

Our ordering is determined by starting with a cycle in $A$, assigning $x$-coordinate
values to the vertices.
Then for each of the vertices in this cycle we select the respective cycles in $B$ and assign
$y$-values.
We now have new vertices that have $y$-values but no $x$-values so we repeat by adding cycles from $A$.

To form our ordering and our embedding, we start by picking an initial cycle $c_0$
from $A$ and an initial vertex $v_0$ in $c_0$.
We pick $c_0$ and $v_0$ so that $c_0$'s vertices belong to more than
one cycle of $B$ and so that $c_0$'s final vertex
belongs to a different cycle of $B$ than $v_0$.
We assign $x$-coordinates to the vertices in $c_0$ appropriately, i.e. in consecutive 
increasing order.

Then, until all vertices have both coordinates assigned, we begin
to assign values to the (unassigned) coordinates.
Among all the vertices with only one assigned coordinate, we choose the one, $v_i$, with the
smallest assigned $y$-coordinate and, if none exist, the smallest assigned $x$-coordinate.
Let the next cycle $c_i$ be the (as yet unassigned) cycle through $v_i$.
We assign the next
available consecutive block of ($x$ or $y$) coordinates to the vertices of $c_i$.
Notice some vertices now will have both coordinates assigned, if they were part of a
previous cycle, while other new vertices will now have one unassigned coordinate instead of two.
We repeat the process by choosing the next vertex and cycle.

Define the {\em back} edge, $e_i$, of a cycle, $c_i$, to be the edge connecting the last vertex $w_i$
in the cycle,
to the first vertex, $v_i$.
After we have embedded our vertices in the above manner, we then add in all edges but
the back edges producing a collection of paths.
Figure~\ref{fig:two-embed} shows an example of our initial embedding.
We claim the following:

\begin{figure}
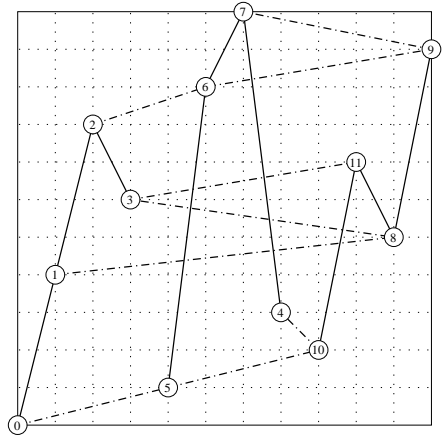

\begin{center}
\ifUsePDF
\includegraphics{two-embed.pdf}
\else
\input{two-embed.pstex_t}
\fi
\end{center}
\vspace{-.5cm}
\caption{\sf\small The initial state of embedding cycles without the final back edges. 
The cycles in $A$ and $B$ are:
$A=\{c_0=\{p_0, p_1, p_2, p_3\}, c_2=\{p_4, p_5, p_6, p_7\}, c_3=\{p_8, p_9, p_{10}, p_{11}\}\}$ and 
$B=\{c_1=\{p_0, p_5, p_{10}, p_4\}, c_4=\{p_1, p_8, p_3, p_{11}\}, c_5=\{p_2, p_6, p_9, p_7\}\}$.
The cycles are labeled, $c_i$, to show their ordering in the embedding.}
\label{fig:two-embed}
\end{figure}

\begin{lemma}
\label{lemma:belowLeft}
After our initial embedding of the cycles and the edges, the first vertex of
each cycle is below and to the left of every other vertex in the same cycle.
\end{lemma}
\begin{proof}
Assume not.
Let $v$ be a vertex that is the first vertex of cycle $c$ but where there is another
vertex $w$ in $c$ that is either below or to the left of $v$.
Without loss of generality, let us assume that $c$ belongs to $A$.
Since $v$ is the first vertex of $c$, $c$ was chosen at the time that $v$ had only one assigned coordinate.
This assignment must have come from a cycle in $B$ implying that the coordinate was a $y$-coordinate value.
Thus, $v$ must have had the lowest $y$-coordinate among all unassigned vertices.
Since $w$ also belongs to $c$, it also could not have had the $x$-coordinate assigned.
If its $y$-coordinate were previously assigned, it would have had to be greater than $v$'s,
and by the construction it will also be assigned an $x$-coordinate greater than $v$'s, contradicting the assumption.
Therefore, neither coordinate of $w$'s could have been assigned, but since successive cycles in $B$
use increasingly larger $y$-coordinates, $w$ will still be both above and to the right of $v$, a contradiction.
\end{proof}

\begin{figure}
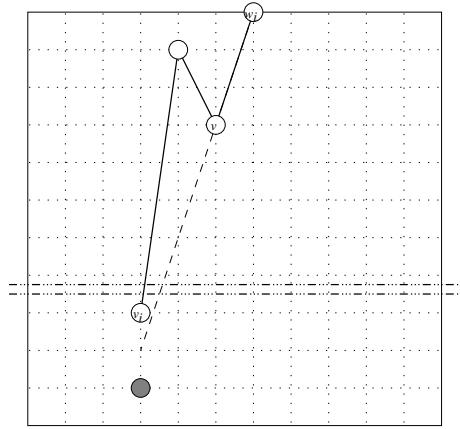

\begin{center}
\ifUsePDF
\includegraphics{insertSpace.pdf}
\else
\input{insertSpace.pstex_t}
\fi
\end{center}
\vspace{-.5cm}
\caption{\sf\small Inserting space directly above $v_i$.
The gray circle indicates a valid position where $v_i$ can see $w_i$ without intersecting any of the cycle edges.}
\label{fig:insertSpace}
\end{figure}

To close the cycles, we now insert the back edges, $e_i=(v_i,w_i)$, 
in the reverse order of the one used to embed the
cycles initially.
That is, we start with the last cycle chosen, $c_k$, and work backwards to $c_0$.
We say vertex $v_i$ {\em sees} $w_i$ if we can connect the two with $e_i$ without
crossing any other edges from the same set.
Since $v_i$ does not necessarily always see $w_i$, at each cycle $c_i$, 
we adjust the grid spacing to accommodate the new back edge.
Specifically, if $c_i$ belongs to $A$, we increase the $y$-coordinate spacing between
$v_i$ and the next unit above until $v_i$ sees $w_i$.
If $c_i$ belongs to $B$, we increase the $x$-coordinate spacing between $v_i$
and the next unit to the right until $v_i$ sees $w_i$.
This process works completely except for the last shared vertex $v_0 = v_1$.
Therefore, we have a few exceptions to the algorithm, which we describe shortly; see Figure~\ref{fig:insertSpace}.

\begin{lemma}
For any cycle $c_i$, after sufficient shifting, $v_i$ will be able to see $w_i$.
\end{lemma}

\begin{proof}
Without loss of generality, assume $c_i$ is in $A$.
From Lemma~\ref{lemma:belowLeft}, we know that $v_i$ is below and to the left of all other vertices
in its cycle.
In addition from the construction, all other vertices and edges in $A$ are either to the left or right of $c_i$.
Therefore, the only obstruction between $v_i$ and $w_i$ is caused by another vertex, $v$, in $c_i$.
More specifically, if we drew a line from $w_i$ through $v$, it would intersect the $x$ position
of $v_i$ at a point $p$ below $v_i$.
Let $v$ be the vertex that forms the lowest such point, $p$.
Increasing the vertical spacing directly above $v_i$ has the same effect
as lowering {\em only} $v_i$ and of course other vertices not in $c_i$.
Therefore, after enough shifts, $v_i$ would be below $p$ and hence able
to see $w_i$; see Figure~\ref{fig:insertSpace}.
\end{proof}

\begin{lemma}
\label{lemma:shiftSafe}
When we insert spacing for a vertex $v_i$ for cycle $c_i$, if $c_j$ is a cycle with $j > i$, then
the relative positions of all vertices and edges in $c_j$ remain fixed.
If $i = 0$, then $c_1$ and its vertex $v_1 = v_0$ can be affected.
\end{lemma}

\begin{proof}
Assume not, let $c_j$ be some cycle where the relative distance between two of its vertices,
$a$ and $b$, changes.
Without loss of generality, let us assume that $c_i$ belongs to $A$.
By our algorithm, we insert spacing above $v_i$.
In order for the distance to change $a$ and $b$ must lie on opposite sides of the 
horizontal line through $v_i$.
That is, $a$ must lie below $v_i$ and $b$ must be above $v_i$
or either $a$ or $b$ actually is the same vertex as $v_i$.

Let us first consider the situation when $c_j$ also belongs to $A$.
Since $j > i$, $v_j$ must be above $v_i$ as when $c_i$ was chosen
it had the smallest $y$-coordinate among all partially assigned vertices.
From Lemma~\ref{lemma:belowLeft}, $a$ and $b$ must also be above $v_i$.

Now, consider the situation when $c_j$ belongs to $B$.
Since $c_j$ comes after $c_i$, when $v_j$ is chosen
$v_i$ has already been assigned both coordinate values.
Since $c_j$ is part of $B$, $v_j$ and the other vertices
in $c_j$ are assigned their $y$-coordinates.
But, these values come from the next available section
of $y$ values which by definition must be larger than $v_i$'s 
$y$-coordinate value.
Therefore, $v_j$ and all other vertices in $c_j$ must lie above
$v_i$, contradicting the definition of $a$ and $b$.

The one exception is the first cycle, $c_i = c_0$.
In this case, for cycle $c_1$, $v_1 = v_0$ and this vertex
is affected as stated in the exception to the lemma.
\end{proof}

\begin{figure}
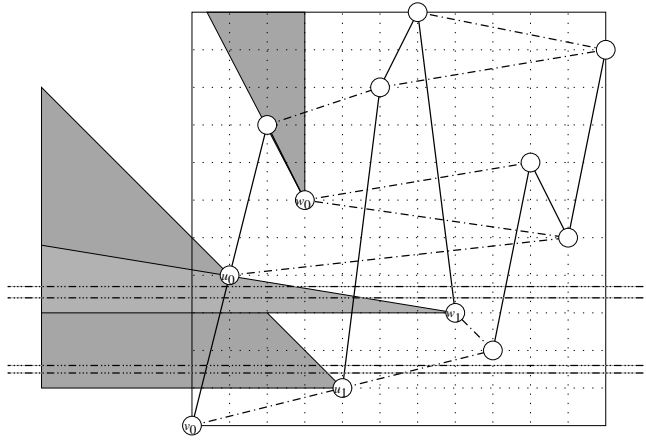

\begin{center}
\ifUsePDF
\includegraphics{wedges.pdf}
\else
\input{wedges.pstex_t}
\fi
\end{center}
\vspace{-.5cm}
\caption{\sf\small A simple example of our wedges defined by vertices $w_0, w_1, u_1$.
Note in this simple example $v_0$ actually sees all four vertices but this
is not always guaranteed.
The dashed lines indicate where the $y$-coordinate shifting is done above $w_1$ and $u_1$
to guarantee that all three wedges will intersect.}
\label{fig:wedges}
\end{figure}

Since we are adding the back edges in reverse order, once a cycle $c_i$ has been fixed, 
any subsequent shifting will no longer affect this cycle and the back edge will remain
properly placed.
However, we must treat the cycles $c_0$ and $c_1$ with special care because
they share the same first vertex $v_0 = v_1$.
Changing one cycle may affect the other.
We need to find a place for $v_0$ which simultaneously sees $w_0$ and $w_1$
as well as its other two neighbors on $c_0$ and $c_1$ which we call
$u_0$ and $u_1$.
For each of these vertices, define a wedge of visibility.
Wedge $W_0$ consists of all those locations from which $v_0$ is high enough
to see $w_0$.
It extends upward from $w_0$.
Wedge $W_1$ consists of all those locations from which $v_0$ sees $w_1$ but is not
high enough to be blocked by the cycle directly above $c_1$.
Wedge $U_0$ consists of those locations from which $v_0$ is high enough to see $u_0$.
Wedge $U_1$ consists of those locations from which $v_0$ see $u_1$ but is not
high enough to be blocked by the cycle.
Since we shall be locating $v_0$ such that it lies inside all four wedges and to the left
of its initial location, we do  not need to consider wedge $U_0$ as $v_0$ will always
lie inside this wedge; see Figure~\ref{fig:wedges}.

As one may observe, it is not always guaranteed that the three wedges intersect.
At first glance, one may wish to use $W_0$ oriented downwards since in our example
this guarantees an intersection.
However, this intersection is not always guaranteed, and we do not see any
simple technique to add spacing that both creates an intersection and does
not affect other cycles in the process.

To guarantee that the intersection of our wedges is non-empty, we must make a few
observations.
First, the initial ordering of cycles meant that $c_0$ was part of $A$, $c_1$
was part of $B$, and then we added possibly several more cycles from $A$,
before adding a second cycle from $B$ which we refer to as $c_b$.
Notice that $c_b$ is the cycle directly above $c_1$ and determines the wedge
formed by $w_1$.
From our initial selection of $c_0$ and $v_0$, we guaranteed that $w_0$ was not
in $c_1$.
This implies that $w_0$ is either in $c_b$ or some even later cycle of $B$.

\begin{lemma}
After the back edge associated with $c_b$ has been added, there will be no
more horizontal spacing insertions until $c_1$'s back edge is added.
\end{lemma}

\begin{proof}
Since horizontal spacing is only added for cycles in $B$, after $c_b$ has been inserted
the only horizontal spacing left is when $c_1$ is visited.
\end{proof}

\begin{lemma}
\label{lemma:fixedWedge}
After the back edge associated with $c_b$ has been added, the wedge $W_0$
defined by $w_0$ is fixed and will not move.
\end{lemma}

\begin{proof}
Vertex $w_0$ belongs to $c_0$ and either $c_b$ or a later cycle.
Once $c_b$ has been shifted, the only shifts that affect it could
come from cycles in $A$ that come after $c_0$ but before $c_b$.
However, these shifts insert only vertical spacing below $w_0$ and its upward
pointing wedge.
Hence, the wedge is unaffected.
\end{proof}

Since there are only vertical shifts after $c_b$ is complete, 
we know that $w_1$'s horizontal position is fixed.
From Lemma~\ref{lemma:fixedWedge}, we know that wedge $W_0$ is also fixed.
Thus, we can actually determine the minimum amount of spacing
needed between $w_1$ and $c_b$ so that $W_1$ intersects $W_0$.
We then insert vertical spacing below $v_b$, the start vertex
associated with $c_b$ until the two wedges intersect.

Guaranteeing that $U_1$ also intersects these two wedges involves a
similar additional spacing shift.
In this case, $u_1$ can be one of two situations.
Either $u_1$ belongs to both $c_0$ and $c_1$ or it is
the first vertex of the cycle in $A$ after $c_0$, which means $u_1$
is part of $c_2$ and $c_1$.
If it is the first vertex of $c_2$, then we simply add more horizontal spacing
directly above $u_1$ to ensure that $u_1 = v_2$ sees $w_2$ and
that wedge $U_1$ also intersects $W_1$ and $W_0$.
Notice that adding space directly above $u_1$ has no effect on wedge $W_1$.
If it is part of $c_0$ and $c_1$, when we reach $c_1$ and begin
to determine $v_1$'s position, we insert vertical spacing above $u_1$ to guarantee
that wedge $U_1$ intersects the other two wedges.
Therefore, at the time that $c_1$ and $c_0$ are ready to be fixed, we have
guaranteed that there exists a position in the plane that can see $u_0$, $u_1$,
$w_0$, and $w_1$, a point inside the intersection of the wedges.
We simply place $v_0 = v_1$ into this location and connect the edges.

This algorithm, the previous lemmas, and the fact that the Euler tour can be computed
in linear time
lead to
the following lemma and subsequent theorem.

\begin{lemma}
Let $A$ and $B$ be two maximum-degree-two graphs on the same vertex set
of size $n$.
A simultaneous geometric embedding of $A$ and $B$
can be found in $O(n)$ time.
\end{lemma}

\begin{theorem}
Graphs with maximum degree four have geometric thickness two.
\end{theorem}

\subsection{An Orthogonal Embedding}

\begin{figure}
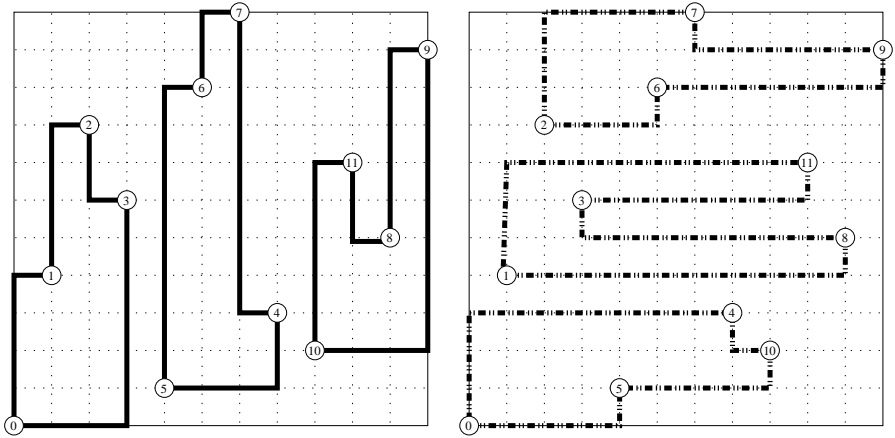

\begin{center}
\ifUsePDF
\includegraphics{ortho.pdf}
\else
\input{ortho.pstex_t}
\fi
\end{center}
\vspace{-.5cm}
\caption{\sf\small An orthogonal two-layer embedding of two forests of cycles.
Since line segments on different layers can overlap, we drew the 
two layers separately.
The cycles are identical to the ones in Figure~\ref{fig:two-embed}.}
\label{fig:ortho}
\end{figure}

The algorithm above uses little area except when it has to adjust the
layout for back edges, particularly back edges involving the initial
two cycles, when we may be forced to have large areas to guarantee
planarity of the two layers.  However, if we instead use orthogonal
edges, we can do better.  Define an orthogonal edge between vertices
$v$ and $w$ to be a sequence of two axis-aligned line segments. Given the two
vertices, there are two possible ways to connect them using orthogonal
edges (starting with a horizontal or vertical line segment from $v$).

Our modification to the algorithm presented in the preceding subsection is
quite simple.
First, we perform the same initial embedding of vertices but avoid the shifts
required of the straight-line algorithm.
Afterwords, we connect edges in the cycle, not counting the back edges, 
using orthogonal edges.
If the cycle is from $A$, we connect vertices using vertical followed by
horizontal line segments.
If the cycle is from $B$, we connect vertices using horizontal followed 
by vertical line segments.
The back edges are done the opposite way;
see Figure~\ref{fig:ortho}.

\begin{theorem}
Graphs with maximum degree four can be embedded in two layers using
orthogonal edges on an $n \times n$ grid in $O(n)$ time.
\end{theorem}

\begin{proof}
From our embedding scheme, we can see that we compute the embedding in linear time, 
that the area is $n \times n$, and
that we use orthogonal edges.
We must then only show that the segments do not intersect.
Without loss of generality, let us look at the segments forming the cycles in $A$.
Recall that each vertex has its own unique $x$ and $y$ coordinate.
In addition, the vertices in cycle $A$ are ordered in increasing $x$-direction.
Therefore the edges that are not back edges will not intersect each other as they
each first are drawn vertically, with no intersection, and then horizontally to the
next vertex, one unit over.
Because each start vertex $v_i$ is below and to the left of all other vertices,
when we connect the back edge, we do not cause any intersection.
The horizontal line from $v_i$ to directly below $w_i$ cannot intersect
any segments and the vertical line segment to $w_i$ can also not intersect
any segments.
The same argument applies for cycles in $B$.
\end{proof}

\section{Implementation}
\label{sec::implementation}
\begin{figure}
\begin{center}
\ifUsePDF
\includegraphics[height=2.5in]{coxeter.pdf}
\qquad
\includegraphics[height=2.5in]{n200.pdf}
\else
\includegraphics[height=2.5in]{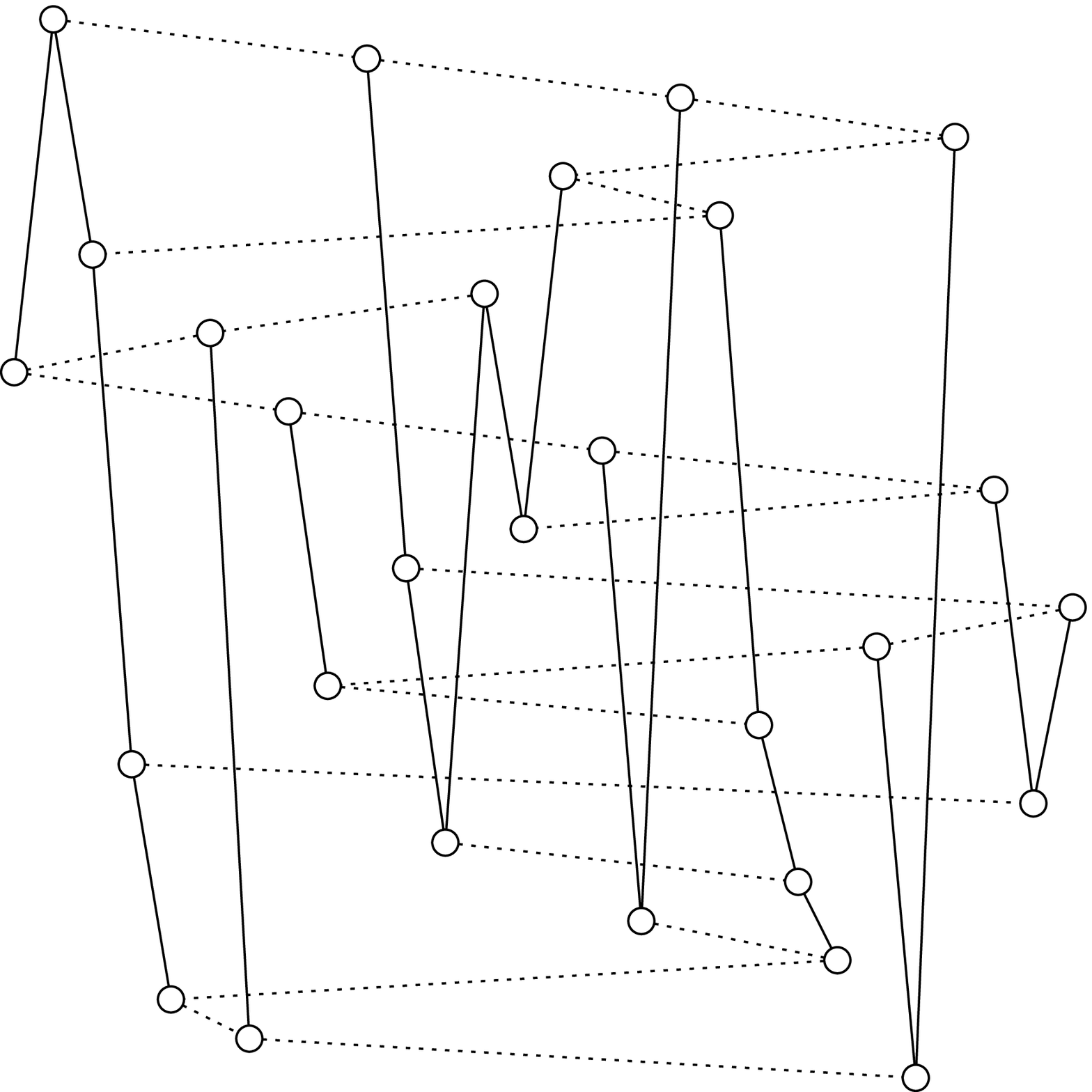}
\qquad
\includegraphics[height=2.5in]{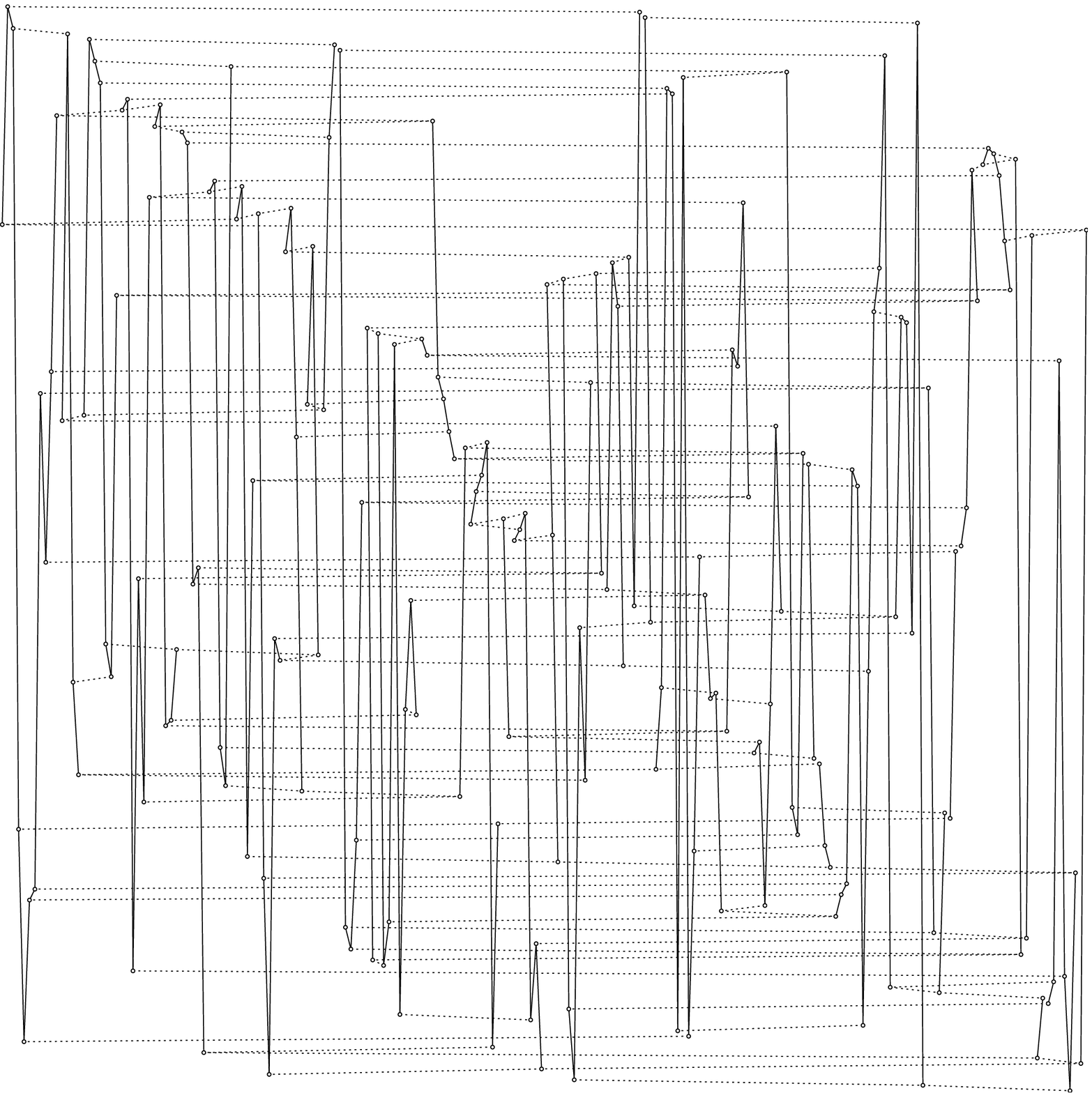}
\fi
\end{center}
\vspace{-.5cm}
\caption{\sf\small A two-layer decomposition of the 3-regular Coxeter graph and of a randomly generated 200-vertex 3-regular graph.}
\label{fig:coxeter}
\end{figure}

We have implemented the algorithms described in
Section~\ref{sec::degreeThree} in Python.  Our implementation reads graphs in various standard formats including MALF, edge list, node list, GraphML, graph6, sparse6, and LEDA-Graph, and
outputs a drawing in SVG format.
Figure~\ref{fig:coxeter} shows
two-layer decompositions produced by our implementation of the Coxeter graph and of a randomly generated 200-vertex graph. The Coxeter graph is a symmetric (vertex-transitive) cubic non-Hamiltonian graph~\cite{Coxeter-graph}.

\section{Conclusion and Open Problems}
\label{sec::conclusion}
We have shown that graphs of maximum degree three and four have
geometric thickness two. The proof of the maximum degree four case
relied on a generalization of the algorithm for simultaneous embedding
of a pair of cycles from~\cite{WADS::Brass2003}. In particular, we
showed that two maximum degree two graphs on the same vertex set can
be simultaneously embedded and the embedding can be found in linear
time.

Several interesting open problems remain:

\begin{enumerate}

\item
Can similar results be obtained for
graphs of maximum degree five?

\item 
Unlike the result for two paths where the required area is $n \times
n$ where $n$ is the size of the vertex set, for the case of two
maximum-degree-two graphs we do not have a good bound on the required grid
area unless we allow a bend in the edges --- can a matching bound be obtained
if we only use straight lines?

\item 
Given two trees on the same vertex set, can they be simultaneously
embedded?

\item
The Coxeter graph example has a lot of inherent symmetry, but
our drawing of it does not.  Can we find low-thickness drawings
of graphs that preserve as much of their symmetry as possible?

\end{enumerate}

\ifSaveSpace
\else
{
\HalfSpace\small
\footnotesize
\raggedright
\bibliographystyle{abbrv}
\bibliography{thickness}
}
\fi

\end{document}